\documentstyle[11pt,amssym,aaspp4]{article}

\begin{document}

\title{SHADES OF BLACK: SEARCHING 
FOR BROWN DWARFS \& GIANT PLANETS}

\author{\sc Gordon Walker, Scott Chapman \& Georgi Mandushev} 
\affil{UBC, Vancouver}

\author{\sc Ren\'e Racine, Daniel Nadeau \& Ren\'e Doyon} 
\affil{U.Montr\'eal}

\author{\sc Jean-Pierre V\'eran}
\affil{DAO, Victoria}

\bigskip
\centerline{}

\bigskip

\begin{abstract} We have searched, so far in vain, for brown dwarfs or giant planets in
the vicinity of several nearby stars using cameras built at Universit\'e de Montr\'eal
attached to the CFHT Adaptive Optics system. Here, we show how pairs of images taken
simultaneously through  separate filters on, and off, the 1.6 $\mu$m methane absorption
band, can be used to completely defeat speckles, the dominant noise source, while
revealing cool brown dwarfs and giant planets in high contrast. We achieve the photon
shot-noise limit for data taken without a focal plane mask but, so far, have had less
success with data where the image is occulted by a mask.
\end{abstract}

\section{THE CHALLENGE}
Figure 1 illustrates the limiting sensitivity needed to detect planetary
companions seen only by reflection like the Earth (E) or Jupiter (J) close to nearby
stars. $\Delta$m is the magnitude difference from the target star, and the
brightness of the planet is assumed to diminish inversely as the square
of angular distance. The abscissa, $a \pi^{-1}$, is the angular
separation, $a$, multiplied by the distance to the system in parsecs,
$\pi^{-1}$. The dashed line to the left of the Jupiter symbol indicates
the locus with changing orbital phase.

%
%
\begin{figure}[ht]
\epsscale{0.6}
\plotone{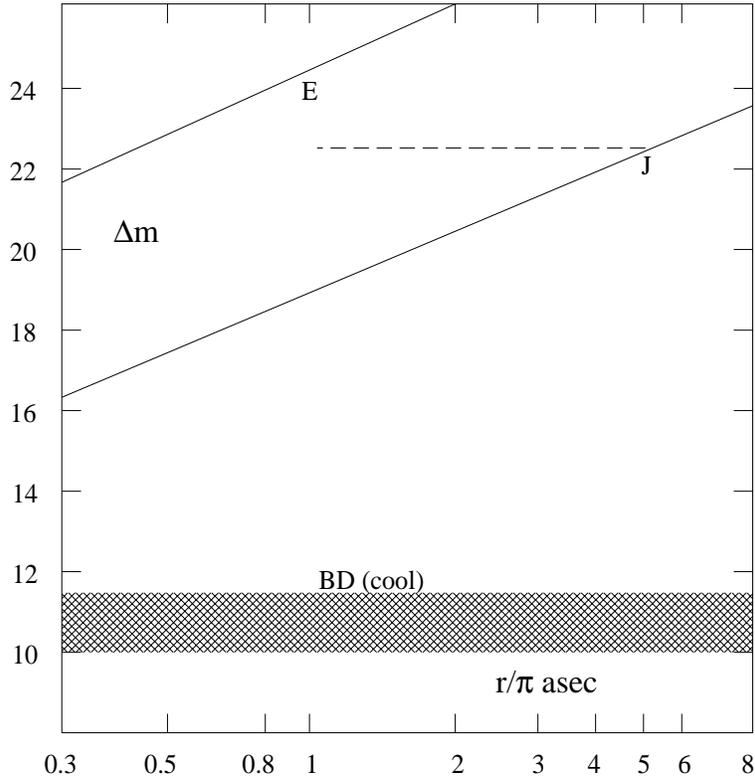}
\caption{The difference in apparent magnitude between a star and planetary 
companions such as the Earth (E) or Jupiter (J) seen only in reflection. The 
lines define an inverse square law at maximum separation. The dashed line 
indicates the effect of changing orbital phase. The $\Delta$m band for brown 
dwarfs covers early M to late-F main sequence stars and  assumes cool brown 
dwarfs have the luminosity of Gliese 229B.}
\end{figure}

The brightness of a cool ($T_{e}\sim1000$) brown dwarf companion like
Gliese 229B, is assumed to be largely independent of separation, and
$\Delta$m depends only on the luminosity of the target star. The band
shown covers roughly the luminosity range for solar-type primaries.
Obviously, cool brown dwarfs should be much easier to detect than
planets, but they appear to be rare indeed with the only convincing
direct image that of Gl229B (\cite{nak95}) and possibly an astrometric
companion (30$M_{J}$, P=530 d) to Gliese 433 detected by Hipparcos and
in speckle images (\cite{net}). The real challenge is to find wide
counterparts to the thirteen short-period Jupiter-mass systems
(\cite{net}) already detected from accurate measurements of their radial
acceleration.

\section{THE SPECKLE PROBLEM}
Figure 2 is a contour plot showing the squared, absolute difference
between two, 0.5 sec exposures at H, taken only seconds apart, of the
bright star 55 Cancri. The data were obtained with the
AO system on CFHT and the MONICA camera (see below) with the image centre
occulted by a focal-plane mask of 0.35 asec radius.. These squared
residuals show the unmatched speckle structure between the two images
in very high contrast. We name the speckles visible in Figure 2,
`super-speckles', because they persist much longer than the few
millisecond dwell-time typical of atmospheric turbulence. The core PSF
of individual speckles appears to be identical to that of the target image. In
Figure 3 we have applied DAOPHOT to the data in Figure 2, treating it as
a globular cluster, and the magnitudes of the `squared' speckles are
displayed as a function of angular distance from the image centre. The
angular distribution of speckle approximates the PSF.

%
%
\begin{figure}[ht]
\epsscale{0.7}
\plotone{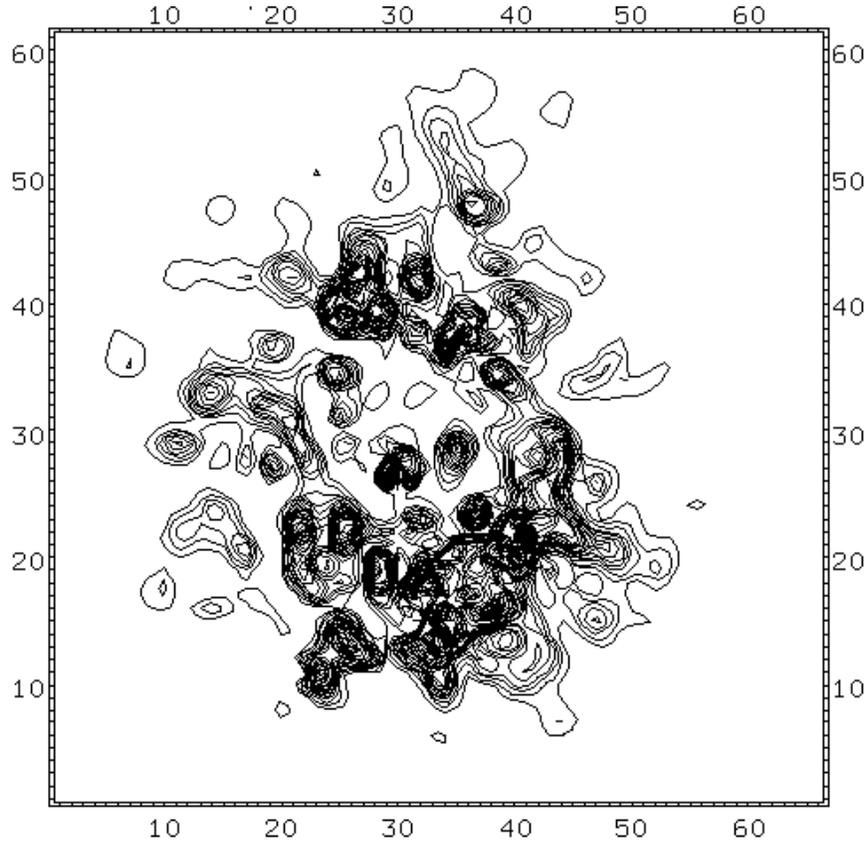}
\caption{This image was formed by subtracting two successive, short (0.5 sec) 
exposures on a bright star (55 Cancri) and squaring the residuals which enhances 
the unmatched speckle pattern between the two images. The image is 2.2 asec on a 
side and the centre was obscured by a focal plane mask of 0.35 asec radius. The 
speckles visible in this figure are longer lived than those expected from 
atmospheric turbulence, we call them `super-speckles', with the brightest defining 
the first diffraction ring.}
\end{figure}

\begin{figure}[ht]
\epsscale{0.75}
\plotone{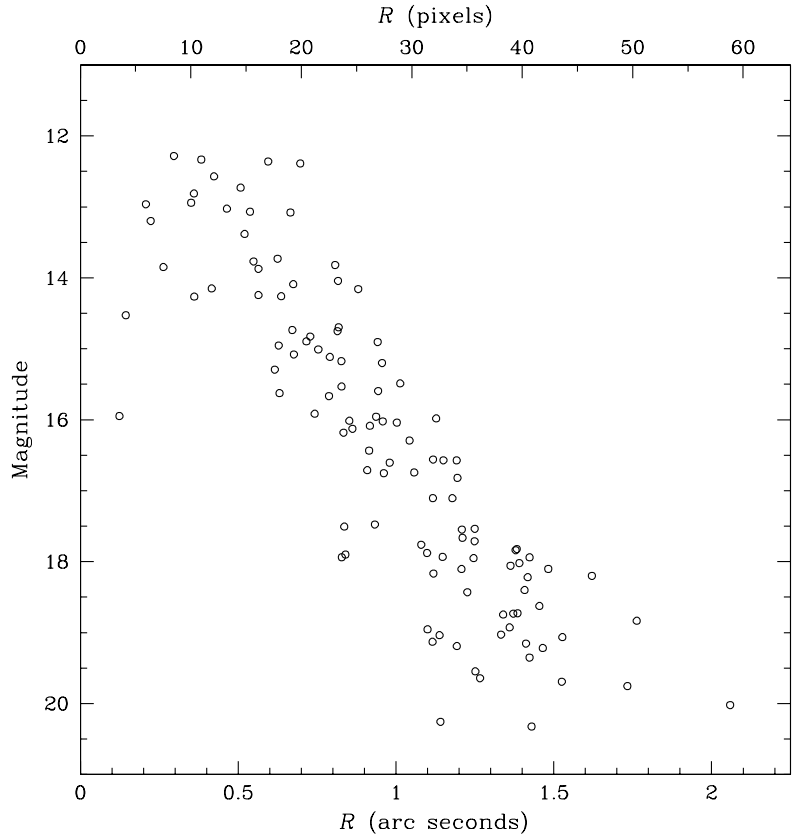}
\caption{The result of treating the speckle pattern in Figure~2 as a globular 
cluster using DAOPHOT. Instrumental magnitudes are shown as a function of 
angular distance form the image centre. The angular intensity (not the squared 
intensity) distribution follows the PSF.}
\end{figure}

Racine presents a full analysis of speckle shot-noise in another paper
at this meeting, pointing out that, because the speckles are identical
to the signal expected from a faint companion, they represent the true
noise limit to detection, making it orders of magnitude brighter than
photon shot-noise for bright stars. In this paper, we describe a
simultaneous dual-imaging technique which appears to successfully
defeat speckle noise.

\section{DUAL-IMAGING} 
%
%
\begin{figure}[ht]
\epsscale{0.80}
\plotone{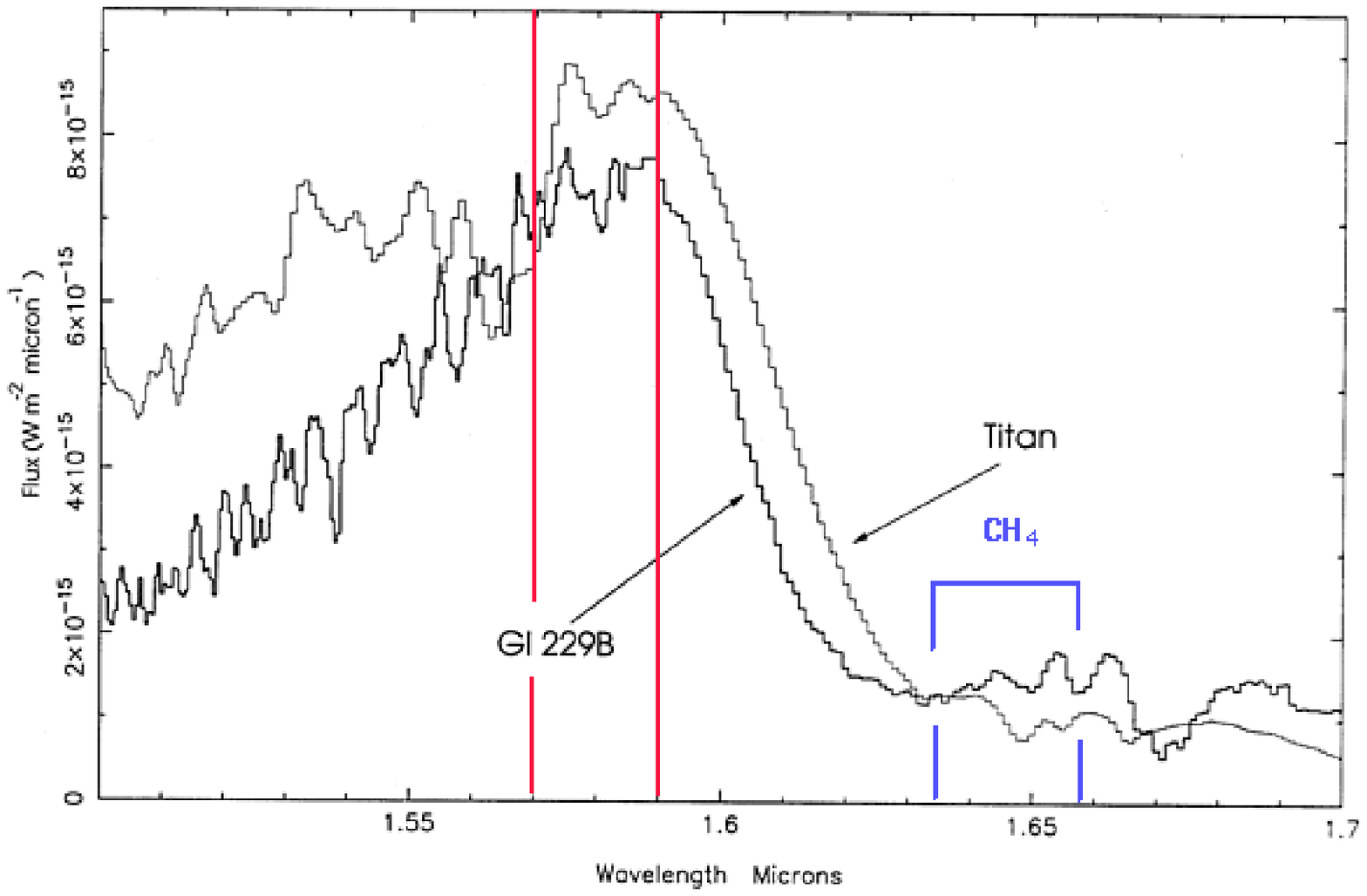}
\caption{The methane 1.6 $\mu$m decrement in spectra of Titan and Gl 229B.
The filters isolate continuum and decrement radiation.}
\end{figure}

%
%
\begin{figure}[ht]
\epsscale{0.20}
\plotone{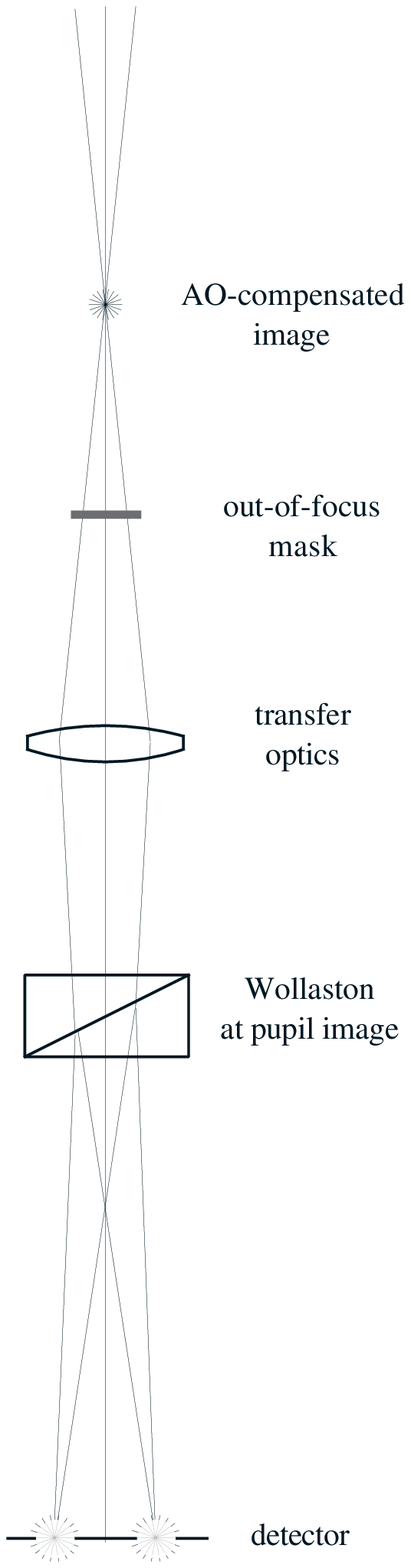}
\caption{A schematic of the optical arrangement used to take simultaneous images 
through the two filters. A filter is placed in each of the emergent beams.}
\end{figure}

Figure 4 shows the methane decrement near 1.6 $\mu$m in spectra of
Titan and Gl 229B (\cite{geb96}). Hotter brown dwarfs and stars display no such
feature and we have selected the two filter bands shown to isolate
the continuum and the decrement. One filter is placed in each of the
beams emerging from the Wollaston prism in the optical arrangement
shown schematically in Figure 5.

Figure 6 shows the `detection' of a brown dwarf using this dual imaging
scheme, where the image was composed by the addition of a G1 229B image to a
field in Orion. The continuum is on the right, the methane decrement
on the left. The continuum image of Gl 229B stands out in sharp
contrast on the right. Giant planets in the Solar system show a similar
decrement to that shown on Figure 4 and we assume that extra-solar
system gas-giants will show a similar contrast to that in Figure 6.

%
%
\begin{figure}[ht]
\epsscale{0.6}
\plotone{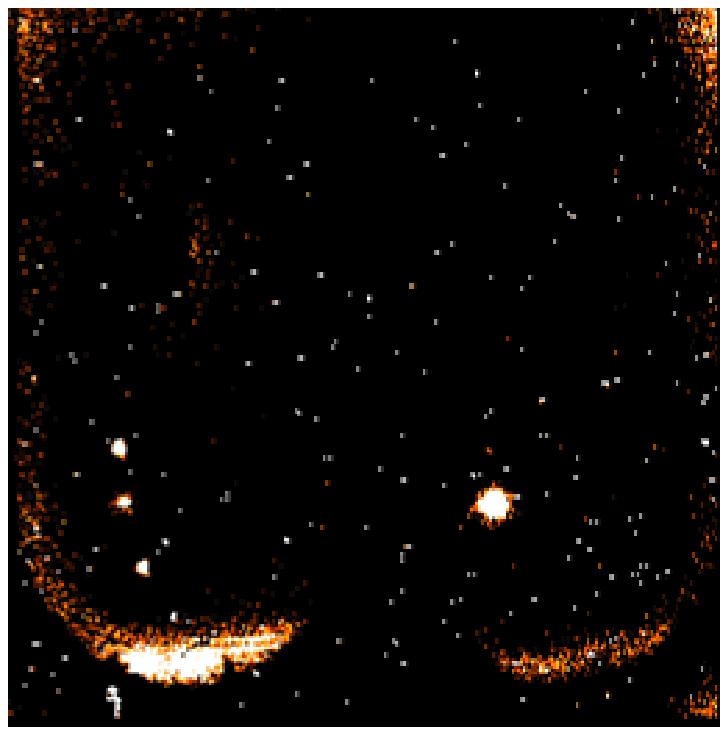}
\caption{A dual image taken with the scheme in Figure 5 on the CFHT AO system. The 
left hand image is through the methane filter, the other in the continuum. The 
image is a combination of a field in Orion and of Gl229B.  The continuum filter 
is the narrower making the Orion stars invisible while Gl229B stands out in high 
contrast.}
\end{figure}

\section{THE DATA}

\vspace{2 mm}
We have had three observing runs with AOB on CFHT, the first two being with the MONICA
camera. A focal plane occulting mask was added for the initial MONICA run, while the 
dual imaging system discussed in the previous section was implemented for the second
run together with an out of focus hard mask. The mask was put out of focus to
provide a `soft' roll-off at the edge of the mask to avoid ringing and improve image
registration. The KIR camera without an occulting mask was used for the third run. Here,
we discuss only the second set of data which used the dual imaging module.

The CFHT AOB system is based on curvature sensing and has a 19-zone
bimorph mirror (\cite{rig98}). MONICA (\cite{nad94}) is a camera built by the Universit\'e
de Montr\'eal group. It has a 256$\times$256 Rockwell NICMOS3 array where
the pixel size corresponds to 0.034 asec pxl$^{-1}$. The data were
obtained on the nights of 1997 December 4 \& 5 when seeing was
$\sim$1 asec and the sky partially clear. The conditions for good AO
imaging were marginal at best.

In order to eliminate the common speckling pattern between the
pairs of images, we reduce the data according to the following scheme:

- bad pixel ($\sim$1500) correction/interpolation

- correlated noise (60 Hz) suppression

- flat-field division (= bright dome flat  $-$ dark dome flat)

- sky subtraction for the un-masked data only

\noindent
the continuum and methane filtered images are then analysed separately:

\begin{center}
\begin{tabular}{l|l|l}
\hspace{11 mm}  methane&~ methane $-$
continuum&\hspace{6 mm}
 continuum\\ 
align images&&align images\\ 
add $ \rightarrow$ long exposure&&add \\ 
&&$\lambda$ scaling for 
diffraction and turbulence\\
&&$ \rightarrow$ long exposure\\
&align long exposures&\\
&scale intensities&\\
&remove background tilt&\\
&subtract $\rightarrow$ residual image &
\end{tabular}
\end{center}

The $\lambda$ scaling for diffraction and turbulence is the most critical step.
The Fourier transform of the infinite integration time, non-speckled, AO image
of a point source at wavelength $\lambda$ can be expressed as (\cite{V97}):
\begin{equation}
B(\lambda,\vec{\rho}) = B_0(\lambda,\vec{\rho})B_s(\lambda,\vec{\rho})
                        \exp \left[ \left(\frac{2\pi}{\lambda}\right)^2
                        D_\phi(\vec{\rho})\right]
\end{equation}
Where $B_0$ is the modulation transfer function (MTF) of an ideal telescope
and can be computed analytically,
$B_s$ is the MTF of the static aberrations within the telescope and
$D_\phi$ is the structure function of $\phi$, the turbulent
wave-front after AO correction, expressed as wavelength independent optical path differences.
If we assume that the finite integration time, speckled, AO image has the
same dependence in wavelength and if we consider that over the
small wavelength range of interest, the telescope static aberrations are
constant, a continuum band ($\lambda_c$) image can be ``translated'' into a
methane band ($\lambda_{m} > \lambda_c$) by the following non-linear
operation in the Fourier domain:
\begin{equation}
B(\lambda_{m},\vec{\rho}) = B(\lambda_{c},\vec{\rho})
             \frac{B_0(\lambda_{m},\vec{\rho})}{B_0(\lambda_{c},\vec{\rho})}
\left| \frac{B(\lambda_{c},\vec{\rho})}{B_0(\lambda_{c},\vec{\rho})} \right|
^{\frac{\lambda_{m}^2}{\lambda_{c}^2}-1}
\end{equation}
where the notation $|a|$ represents the modulus of any complex number $a$.
To be successful though, this operation requires the image to be perfectly
centered, which is a problem with the masked data. Also, special care
must be taken to allow the high spatial frequency component (past the
telescope cut-off frequency) of the noise to be preserved and in order
to avoid ringing effects.

\section{ESTIMATING SHADES OF BLACK}
We estimate limiting magnitudes as follows:

\begin{description}
\item - correlate image with a truncated PSF probe which = 1 when correlated with the
target star image,

\item - generate a circular average from the correlation image.
\end{description}

The results of this analysis are summarised in Figure 7 for a total
exposure of 1000 s on 55 Cancri (G8 V, m$_{H}$=4.4), with no focal
plane mask. The upper, solid curve corresponds to the $\Delta$m
achieved with speckle suppression by the full subtraction and scaling
scheme outlined in section 4, the lower solid line corresponds to the limit in
the continuum filter alone without any speckle correction. The improvement of
one to two magnitudes in sensitivity represents the gain through speckle suppression
for this star and this exposure time. The results tally well with those
predicted by Racine who finds close agreement between his model and our
results.

%
%
\begin{figure}[ht]
\epsscale{0.65}
\plotone{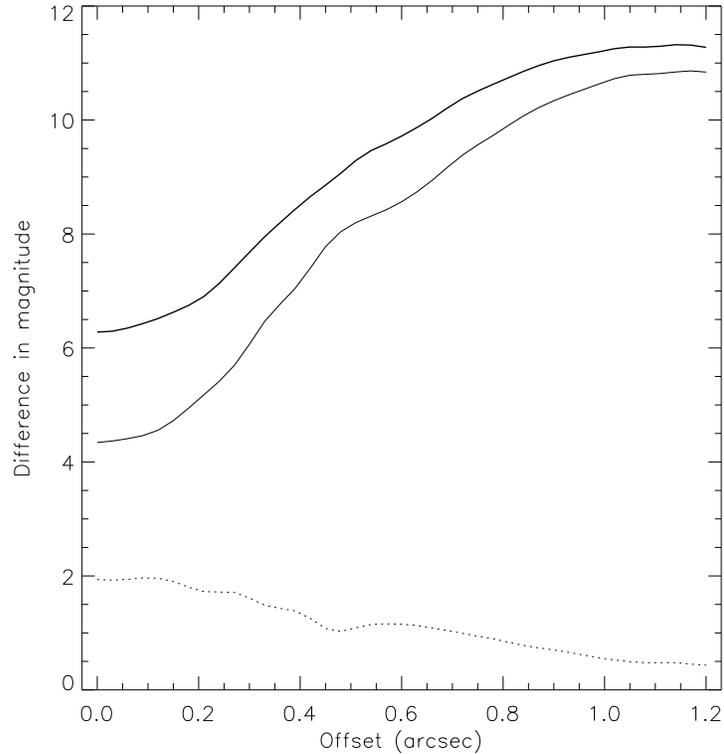}
\caption{Limiting magnitudes for companion detection in 1000 s for the m$_{H}$=4.4, 
G8 V, star 55 Cancri. The upper thick, solid line is the limit with speckle 
suppression, the lower, thin line is for the continuum filter alone with no 
speckle compensation. The dotted line is the difference between the two solid 
curves indicating the increasing advantage of speckle supression towards the 
image centre.}
\end{figure}

The great challenge is to push for the deepest shades of black in the
glare close to the target star when the centre of the image is occulted
by a mask excluding $>$90\% of the star light. The principal advantage
in our case is the lack of detector saturation as we did not have a
true coronagraphic configuration with a pupil-plane mask. To date, we
have been less successful in speckle cancellation with the masked
images, probably because of several problems peculiar to this data set.
There was a small differential parallax between the mask and the two
images caused by the mask being out of focus. We corrected for the
parallax by imposing an enlarged numerical mask before alignment. The
alignment is less accurate than for the unmasked images where the image
centre is well defined. Our analysis will continue.

The current AO system plus infrared cameras is also plagued by several
prominent ghost images, some close to the main image, not all of which
are common or of the same relative intensity in both images.  Again,
we  excluded some ghosts with numerical masks.  The Rockwell NICMOS3
array in the MONICA camera has severe persistence which makes
flat-fielding a challenge in the brightest regions of the image, and
can easily lead to faint companion `discoveries' when the target star
image dwells in an off-mask position for any time during set-up. We
consider that all of these are soluble problems with careful optical
design and a new generation of detectors.

\end{document}